\documentclass[aps,prb,notitlepage,twocolumn,longbibliography,nofootinbib]{revtex4-2}

\usepackage{amsmath,amssymb,amsthm,bm}
\usepackage{graphicx}
\usepackage[colorinlistoftodos]{todonotes}
\usepackage[inline]{enumitem}
\definecolor{darkblue}{rgb}{0.1,0.2,0.6}
\definecolor{darkred}{rgb}{0.8,0.1,0.2}
\definecolor{crimson}{RGB}{164,16,52}
\definecolor{darkgreen}{rgb}{0.31,0.62,0.24}
\usepackage[colorlinks=true, allcolors=crimson]{hyperref}
\usepackage{diagbox}
\usepackage{float}
\usepackage{multirow}
\usepackage{tikz}
\usepackage{braket}
\usepackage[all]{xy}
\usepackage{tikzsymbols}

\usepackage{bbm}

\newcommand{\Rom}[1]{\uppercase\expandafter{\romannumeral#1}}

\newcommand{\da}{\dagger}

\newcommand{\up}{\uparrow}
\newcommand{\down}{\downarrow}

\newcommand{\mc}{\mathcal}

\newcommand{\ZZ}{\mathbb{Z}}

\DeclareMathOperator{\Tr}{Tr}

\newtheorem*{claim*}{Claim}

\newcommand{\dia}[3]{\raisebox{#2pt}{\includegraphics{dia_#1}}\hspace{#3pt}}

\usepackage[normalem]{ulem}

\newcommand{\comment}[1]{}

\begin{document}
\title{Efficient Preparation of Nonabelian Topological Orders in the Doubled Hilbert Space}
\author{Shang Liu}
\email{sliu.phys{\Laughey}gmail.com}
\affiliation{Kavli Institute for Theoretical Physics, University of California, Santa Barbara, California 93106, USA}

\begin{abstract}
	Realizing nonabelian topological orders and their anyon excitations is an esteemed objective. In this work, we propose a novel approach towards this goal: quantum simulating topological orders in the \textit{doubled Hilbert space} -- the space of density matrices. We show that ground states of all quantum double models (toric code being the simplest example) can be \emph{efficiently} prepared in the doubled Hilbert space; only finite-depth local operations are needed. In contrast, this is not the case in the conventional Hilbert space: Ground states of only some of these models are known to be efficiently preparable. Additionally, we find that nontrivial anyon braiding effects, both abelian and nonabelian, can be realized in the doubled Hilbert space, although the intrinsic nature of density matrices restricts possible excitations. 
\end{abstract}

\maketitle

\section{Introduction}
In this Noisy Intermediate-Scale Quantum (NISQ) era, quantum simulation has attracted significant amount of research efforts \cite{QSimuRoadmapPRXQuantum,QSimuNature2022}. One major question being actively explored is how to realize certain useful or novel quantum states on a realistic platform.  

In the pursuit of this endeavor, it has been realized that \emph{nonunitary operations} are extremely useful tools in many quantum simulation tasks. 
Indeed, as shown in Refs.\,\onlinecite{Nat2021KWfromMeasurement,Nat2021NonabelianTO,Nat2023Shortest,Nat2023Hierarchy,Lu2022Measurements}, certain long-range entangled many-body quantum states
can be prepared with constant layers of local unitary gates and local measurements (see also Refs.\,\onlinecite{Raussendorf2001Cluster,Raussendorf2005NoisyCluster}). In contrast, if only local unitaries are allowed, one would need a unitary circuit whose depth increases with the system size. However, although quantum measurements are efficient for generating long-range entanglement, they have one obvious shortcoming -- the randomness of outcomes. In a limited set of cases discussed in the references above, it is possible to efficiently cure the randomness from measurements. In other situations, it seems that one has to do inefficient post-selections or seek alternative protocols, unless when an intrinsically random state is wanted \cite{Ji2022Measurements}. 

In this work, we propose a novel route of quantum simulation where \emph{deterministic} nonunitary operations are possible: quantum simulation in the \emph{doubled Hilbert space} -- the space of density matrices. More explicitly, each density matrix $\rho$ of a quantum state can be mapped to a pure state $\ket{\rho}$ in the doubled space. The wave function of $\ket{\rho}$ is nothing but the density matrix elements $\rho_{ij}$. 
Applying a quantum channel on $\rho$ is equivalent to acting a generically nonunitary operator on $\ket{\rho}$. Therefore, it is possible that interesting quantum states hard to prepare in the conventional Hilbert space can be efficiently prepared as states $\ket{\rho}$ in the doubled Hilbert space. 


We show that this is indeed the case through the example of realizing topologically ordered states, which is perhaps the most exciting subfield of quantum simulation due to its direct relevance to fault-tolerant quantum computation \cite{TC_SC_2021,TC_Rydberg_2021,Google2023NonabelianAnyon,Dreyer2023TC,Kim2023AdaptiveCircuits,Dreyer2023D4}\cite{Rhine2021KagomeRydberg,Ruben2021RubyRydberg,Nat2021KWfromMeasurement,Nat2021NonabelianTO,Nat2023Shortest,Nat2023Hierarchy,BaoFan2023TODecoherence1,BaoFan2023TODecoherence2}. More specifically, we show that in the doubled Hilbert space, with constant-depth local operations, 
one can prepare ground states of Kitaev's quantum double models \cite{Kitaev2003ToricCode}, or $G$-gauge theories, for all finite groups $G$. In the conventional Hilbert space, however, the same task is known to be possible only for a subset of groups (solvable groups) \cite{Nat2021KWfromMeasurement,Nat2021NonabelianTO,Nat2023Hierarchy}. 

This is not a free lunch though. The advantage of doubled Hilbert space comes with a cost: Intrinsic properties of density matrices restrict the allowed states and not all excited states of these topological orders are accessible. Nonetheless, we show that nontrivial anyon braiding, a hallmark of topological order, can be realized in the doubled Hilbert space. This is true for both abelian and nonabelian braiding. 

Doubled-space topological order also explicitly or implicitly appears in a few previous works \cite{JYL2023CriticalityDecoherence,BaoFan2023TODecoherence1,Liu2023QuantumDims,Wang2023SteadyStateTO1,Wang2023IntrinsicMixedStateTO,Wang2023SteadyStateTO2}, but they are all quite different from the present one. 
More specifically, Refs.\,\onlinecite{JYL2023CriticalityDecoherence,BaoFan2023TODecoherence1,Liu2023QuantumDims,Wang2023IntrinsicMixedStateTO} assume a topologically ordered state already exists in the conventional Hilbert space, while in this work, we begin with a product state. Refs.\,\onlinecite{Wang2023SteadyStateTO1, Wang2023SteadyStateTO2} consider abelian doubled-space topological orders as steady states of dissipative systems, while in this work, we prepare both abelian and nonabelian doubled-space topological orders by few-layer local quantum channels. 

In the following, we will start from an abelian example: building a toric code ground state in the doubled Hilbert space and realizing abelian anyon braiding. We then discuss more general quantum doubled models and the phenomenon of nonabelian anyon braiding.

\section{Example of Toric Code}
Given a many-body density matrix $\rho$ and a real-space tensor product basis $\{ \ket{i} \}$ of the Hilbert space, we can write the density matrix as $\rho=\sum_{i,j}\rho_{ij}\ket{i}\bra{j}$, and then map it to a pure state in the doubled Hilbert space: 
\begin{align}
	\ket{\rho}&:=\frac{1}{\sqrt{\Tr(\rho^2)}}\sum_{i,j}\rho_{ij}\ket{i}\ket{j}. 
\end{align}
We will refer to these two copies of Hilbert spaces as the ket and bra spaces, respectively. 

In this paper, we will introduce a strategy for building up topologically ordered states in the doubled Hilbert space. In this section, we will first consider the simplest example -- that of the toric code model \cite{Kitaev2003ToricCode}. More complicated topological orders, especially nonabelian ones, will be discussed in later sections. 
Given a square lattice with qubits living on the edges/links, the toric code model is defined by the following Hamiltonian 
\begin{align}
H_{\rm TC}&=-\sum_v\dia{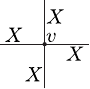}{-19}{0}-\sum_f\dia{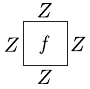}{-19}{0}, \nonumber\\
&\equiv -\sum_v\mc {X}_v-\sum_f\mc {Z}_f. 
\label{eq:TCHamiltonian} 
\end{align}
Throughout this paper, we use $v$ ($e$, $f$) to label vertices (edges, faces). $\mc X_v$ ($\mc Z_f$) is the product of all Pauli-$X$ (Pauli-$Z$) operators on the neighboring edges of vertex $v$ (face $f$). Generalizations of the Hamiltonian to non-square lattices are straightforward. For simplicity, we can assume the lattice to be periodic, although open boundary conditions are also not hard to analyze. $H_{\rm TC}$ is a frustration free Hamiltonian: Ground states of it satisfies each single term, i.e. $\mc X_v=\mc Z_f=1$ for all $v,f$. 

Let us now describe our protocol for building a toric code ground state in the doubled Hilbert space. We again consider a periodic square lattice with qubits living on edges/links, and take the initial state $\ket{0}$ to be the product state $\prod_e\ket{\up}_e$. The initial density matrix is denoted as $\rho_0=\ket{0}\bra{0}$, and the corresponding state in the doubled space is $\ket{\rho_0}=\prod_{e,e'}\ket{\up}_e\ket{\up}_{e'}$. 
For each vertex $v$ on the lattice, we define a local quantum channel $\mc E_v[\cdot]$ which acts on an arbitrary density matrix $\sigma$ as 
\begin{align}
\mc E_v[\sigma]=\frac{1}{2}(\sigma+\mc X_v \sigma\mc X_v). 
\end{align}
Let $\mc E:=\prod_v \mc E_v$ be the composition of all $\mc E_v$; the order does not matter because the $\mc X_v$ operators all commute. The next step of our protocol is to take $\rho_0$ through the quantum channel $\mc E$. Due to the locality of $\mc X_v$, this quantum channel can be realized by applying a finite-depth local unitary circuit on $\ket{0}$ together with some auxiliary spins. 

\begin{figure}
	\centering
	\includegraphics{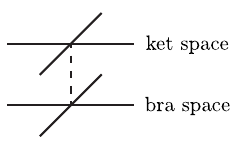}
	\caption{Illustraction of the doubled lattice. }
	\label{fig:DoubledLattice}
\end{figure}

Let $\rho$ be the resulting density matrix, and let $\ket{\rho}$ be the corresponding pure state in the doubled space. We have 
\begin{align}
\ket{\rho}\propto\prod_v\left[\frac{1}{2}(1+\mc X_v\otimes \mc X_v) \right]\ket{0}.  
\end{align}
We claim and will now explain that $\ket{\rho}$ has the toric code topological order. We may regard $\ket{\rho}$ as living on two copies of the original lattice, as illustrated in Fig.\,\ref{fig:DoubledLattice}. 
For each edge $e$ (vertex $v$, face $f$) in the ket space lattice, we denote its counterpart in the bra space lattice by $\bar e$ ($\bar v$, $\bar f$). 
With this notation, we observe that $\ket{\rho}$ has $+1$ eigenvalue under the following mutually commuting operators/stabilizers: 
\begin{itemize}
	\item $Z_eZ_{\bar e}$ for all $e$, 
	\item $\mc Z_f$ and $\mc Z_{\bar f}$ for all $f$. 
	\item $\mc X_v\mc X_{\bar v}$ for all $v$. 
\end{itemize}
The first set of stabilizers implies that each pair of edges $e$ and $\bar e$ can be identified in the $Z$-basis. Under this identification, the bilayer lattice reduces to a single layer, and the third set of stabilizers is equivalent to $\mc X_v$. Combining with the second set of stabilizers, we see that $\ket{\rho}$ indeed minimizes the toric code Hamiltonian $H_{\rm TC}$. There is also an alternative perspective: Just imagine gluing each pair of vertices $v$ and $\bar v$ as shown by the dashed line in Fig.\,\ref{fig:DoubledLattice}. We obtain a lattice where each pair of (new) vertices are connected by two edges. The stabilizers listed above imply that $\ket{\rho}$ minimizes the toric code Hamiltonian defined on this new lattice.

\section{Restricted Excitations}
The instrinsic properties of density matrices, including Hermiticity, positivity, and trace unity, constraint possible operations in the doubled Hilbert space. As a consequence, the allowed excitations above the aforementioned ground state in the doubled space are restricted. For example, $\Tr(\rho)=1$ implies that $\ket{\rho}$ has a nontrivial overlap with the state $\ket{I}:=D^{-1/2}\sum_{i}\ket{i}\ket{i}$, where $\{\ket{i}|i=1,2,\cdots,D\}$ is an orthonormal basis of the $D$-dimensional conventional Hilbert space. Since $\ket{I}$ satisfies (has $+1$ eigenvalue under) the first and third sets of stabilizers mentioned previously, i.e. $Z_eZ_{\bar e}$ and $\mc X_v\mc X_{\bar v}$, no matter what operations we do on $\ket{\rho}$, the resulting state has no hope to completely violate (have $-1$ eigenvalue under) these stabilizers. 


\section{Abelian Anyon Braiding}
Although the accessible excitations are restricted as explained above, we will show that nontrivial anyon braiding, a hallmark of topological order, can still be implemented in the doubled Hilbert space. 

\begin{figure}
	\centering
	\includegraphics{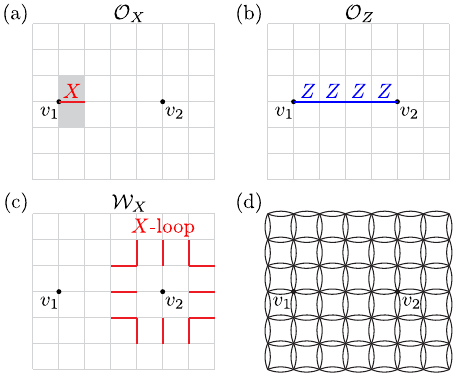}
	\caption{(a-c) Definition of the $\mc O_X$, $\mc O_Z$, and $\mc W_X$ operators used for demonstrating abelian braiding. (d) One view of the doubled Hilbert space. }
	\label{fig:AbelianBraiding}
\end{figure}
Let $\rho$ be the density matrix that has toric code topological order in the doubled Hilbert space. To describe our protocol, define $\mc O_X$, $\mc O_Z$ and $\mc W_X$ operators as shown in Fig.\,\ref{fig:AbelianBraiding}a-c. Also define $U=(\mc O_X+\mc O_Z)/\sqrt{2}$. We propose comparing the following two states in the doubled space. 
\begin{enumerate}
	\item $(\mc W_X \otimes \mc W_X )\ket{\rho}=\ket{\rho}$. 
	\item $(U\otimes U)(\mc W_X \otimes \mc W_X )(U\otimes U)\ket{\rho}=: \ket{\rho'}$. 
\end{enumerate}
Here we have used the fact that $U$ and $\mc W_X$ are both real unitary operators. The $U\otimes U$ operator does not respect the $Z_eZ_{\bar e}=1$ constraints, hence we can no longer identify the edges from the two lattice layers. We can still use the alternative picture mentioned previously: Only identify the vertices from the two layers, and obtain a lattice of the form in Fig.\,\ref{fig:AbelianBraiding}d. 
$U\otimes U$ creates excitations on this lattice, and in particular, there is a chance that a charge (a vertex with $\mc X=-1$) is generated at $v_2$. 
$\mc W_X\otimes W_X$ creates a pair of fluxes (faces with $\mc Z=-1$) from the vacuum, winds one flux around $v_2$, and annihilates the flux pair back to the vacuum. 

To see how $\ket{\rho'}$ is related to anyon braiding, we write 
\begin{align}
	&(U\otimes U)\ket{\rho}=\ket{\rho_+}+\ket{\rho_-},\\
	&\ket{\rho_+}=\frac{1}{2}(1+\mc O_X\otimes \mc O_X)\ket{\rho},\\
	&\ket{\rho_-}=\frac{1}{2}(\mc O_X\otimes\mc O_Z+\mc O_Z\otimes \mc O_X)\ket{\rho}, 
\end{align}
where $(\mc O_Z\otimes \mc O_Z)\ket{\rho}=\ket{\rho}$ has been used. $\ket{\rho_+}$ contains no excitation near $v_2$, but $\ket{\rho_-}$ contains a charge at $v_2$. After braiding a flux, we have
\begin{align}
	(\mc W_X\otimes \mc W_X)(U\otimes U)\ket{\rho}=\ket{\rho_+}-\ket{\rho_-}, 
\end{align}
where the minus sign is the result of nontrivial (semionic) mutual statistics between a charge and a flux. Finally applying $(U\otimes U)$ again, we find
\begin{align}
	\ket{\rho'}&=(U\otimes U)(\mc W_X\otimes \mc W_X)(U\otimes U)\ket{\rho}\nonumber\\
	&=(\mc O_X\otimes \mc O_X)\ket{\rho}, 
\end{align}
which contains a pair of separated magnetic fluxes and thus differs from $\ket{\rho}$. 

The difference between $\rho$ and $\rho'$ can be detected using the following method: Prepare an ancillary qubit in the $\ket{\up}$ state, and then send the whole system $(\ket{\up}\bra{\up})\otimes \rho'$ through the following quantum channel. 
\begin{align}
	\mc E_f[\sigma]&:=[Z\otimes (1+\mc Z_f)/2]\sigma[Z\otimes (1+\mc Z_f)/2]\nonumber\\
	&+[X\otimes (1-\mc Z_f)/2]\sigma[X\otimes (1-\mc Z_f)/2], 
\end{align}
where $f$ is one of the two shaded faces in Fig.\,\ref{fig:AbelianBraiding}a. The resulting ancillary qubit is in the $\ket{\down}$ state. In contrast, if $(\ket{\up}\bra{\up})\otimes \rho$ is sent through $\mc E_f$, the ancillary qubit will still be in the $\ket{\up}$ state. This concludes our example of abelian braiding. 

We note that although the operator $U$ has a nonlocal nature, it can be step-by-step built up using local unitaries. Let $U_{n}:=(X_1+Z_1Z_2\cdots Z_n)/\sqrt{2}$ be the $n$-qubit version of this operator. $U_1$ is the single-qubit Hadamard gate. General $U_n$ can be constructed by the recursive relation $U_n=\mathrm{CNOT}_{n,n-1}U_{n-1}\mathrm{CNOT}_{n,n-1}$. Here, $\mathrm{CNOT}_{i,j}:=(-1)^{(1-Z_i)(1-X_j)/4}$ is a two-qubit gate and one can check that under the conjugate of this gate, $(Z_j,X_j)\mapsto (Z_iZ_j,X_j)$. 

\section{General Quantum Double Models}
Toric code is the simplest instance of the more general \emph{quantum doubled models} \cite{Kitaev2003ToricCode}. In this section, we will show that ground states of all these models admit efficient preparation in the doubled Hilbert space. This is in sharp contrast with the case of conventional Hilbert space, where ground states of only a subset of these models are known to be efficiently preparable \cite{Nat2021KWfromMeasurement,Nat2021NonabelianTO,Nat2023Hierarchy}. 

There is a quantum double model for each finite group $G$ which is generally nonabelian. The model can live on an arbitrary lattice on an arbitrary orientable two-dimensional surface. Physical degrees of freedom, called spins, live on the edges, and the local Hilbert space of each spin is spanned by the orthonormal group element basis $\{ \ket{g}|g\in G \}$. We choose a direction for each edge. Reversing the direction of a particular edge is equivalent to the basis change $\ket{z}\mapsto\ket{z^{-1}}$ for the corresponding spin. Let $v$ be a vertex, and $f$ be an adjacent face, we define the local gauge transformations $A_v(g)$ and magnetic flux operators $B_{(v,f)}(h)$ as follows. 
\begin{align}
	&A_v(g)\Bigg|{\dia{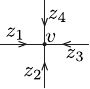}{-19}{0}}\Bigg\rangle=\Bigg|{\dia{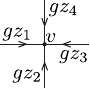}{-19}{0}}\Bigg\rangle,\\
	&B_{(v,f)}(h)\Bigg|{\dia{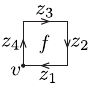}{-19}{0}}\Bigg\rangle=\delta_{z_1z_2z_3z_4,h}\Bigg|{\dia{FaceState}{-19}{0}}\Bigg\rangle. 
\end{align}
Here we use a tetravalent vertex and a square face as examples, and the generalizations should be straightforward. 
We further define two projectors:
\begin{align}
	A_v:=|G|^{-1}\sum_{g\in G}A_v(g),\quad  B_f:=B_{(v,f)}(1). 
\end{align}
Note that $B_f$ does not depend on the choice of the adjacent vertex $v$. The quantum double Hamiltonian is then given by \cite{Kitaev2003ToricCode}
\begin{align}
    H_{\rm QD}=-\sum_v A_v-\sum_f B_f. 
    \label{eq:HQD}
\end{align}
The projectors in $H_{\rm QD}$ all commute with each other, and ground states of $H_{\rm QD}$ satisfy $A_v=B_f=1$. In the special case $G={Z}_2$, if we identify $\ket{\up}$ ($\ket{\down}$) with the group element states $\ket{1}$ ($\ket{-1}$), then $H_{\rm QD}$ is equivalent to $H_{\rm TC}$ in Eq.\,\ref{eq:TCHamiltonian} up to a positive factor and an additive constant. 

The procedure of preparing a ground state of $H_{\rm QD}$ in the doubled space is similar to the previous toric code example. 
Consider an arbitrary orientable two-dimensional lattice with group element states living on the edges/links. We take the initial state $\ket{0}$ to be the product state $\prod_e\ket{1}_e$, where $1\in G$ denotes the identity group element. For each vertex $v$ on the lattice, we define a local quantum channel $\mc E_v[\cdot]$ by
\begin{align}
    \mc E_v[\sigma]=\frac{1}{|G|}\sum_{g\in G} A_v(g) \sigma A_v^\da (g). 
\end{align}
Notice that in the group element basis, each $A_v(g)$ is a real unitary matrix, thus $|G|^{-1}\sum_{g\in G} A_v(g)^\da A_v(g)=1$, and the above map is indeed a valid trace-preserving quantum channel. Let $\mc E:=\prod_v \mc E_v$; the order does not matter because the $A_v(g)$ operators at different vertices commute with each other. The next step of our protocol is to take the initial density matrix through $\mc E$. 
Let $\rho$ be the resulting density matrix, and let $\ket{\rho}$ be the corresponding pure state in the doubled space. We have 
\begin{align}
    \ket{\rho}\propto\prod_v\left[ |G|^{-1}\sum_g A_v(g)\otimes A_v(g) \right]\ket{0}. 
\end{align}
As before, we may regard $\ket{\rho}$ as living on two copies of the original lattice; see Fig.\,\ref{fig:DoubledLattice}. We observe that $\ket{\rho}$ has $+1$ eigenvalue under the following mutually commuting projectors: 
\begin{itemize}
    \item $\sum_{g\in G}\ket{g}_e\ket{g}_{\bar e}\bra{g}_e\bra{g}_{\bar e}$ for all $e$, 
    \item $B_f$ (ket space) and $B_{\bar f}$ (bra space) for all $f$. 
    \item $|G|^{-1}\sum_g A_v(g) A_{\bar v}(g)$ for all $v$. 
\end{itemize}
As in the previous example, there are two ways of seeing the topological order of $\ket{\rho}$: (1) Using the first set of projectors, we can identify each pair of edges $e$ and $\bar e$ in the group element basis, thus reducing the bilayer lattice to a single layer. $\ket{\rho}$ is a ground state of $H_{\rm QD}$ on this reduced lattice. (2) Just identify each pair of vertices $v$ and $\bar v$ as shown by the dashed line in Fig.\,\ref{fig:DoubledLattice}. $\ket{\rho}$ is also a ground state of $H_{\rm QD}$ on this lattice with doubled edges.

One may wonder what is the purification of $\rho$. In other words, how to actually implement the quantum channel $\mc E$? Let us start from the local quantum channel $\mc E_v$. One way to implement this channel is to add an auxiliary group-element spin to the vertex $v$, then prepare this spin in the state $\ket{+}:=|G|^{-1/2}\sum_{g\in G}\ket{g}$, 
and finally apply the following local unitary operator.  
\begin{align}
    U_v:=\sum_{g\in G}\ket{g}_v\bra{g}_v\otimes A_v(g), 
\end{align}
which acts on both the vertex and the surrounding edges. One may check that for an arbitrary density matrix $\sigma$ living on the edges, 
\begin{align}
    \Tr_v\left[ U(\ket{+}_v\bra{+}_v\otimes\sigma)U^\da \right]=\mc E_v[\sigma], 
\end{align}
therefore realizing the desired local quantum channel. Now, to implement the total quantum channel $\mc E$, one just need to add auxiliary spins to all vertices, all prepared in the $\ket{+}$ state, and then apply the \emph{finite-depth} local unitary circuit $U=\prod_v U_v$. Going back to our state preparation protocol, this means that $\rho$ can be obtained from the pure state $\ket{\psi}:=U[(\prod_v\ket{+}_v)\otimes \ket{0}]$ by tracing out all vertex spins. The state $\ket{\psi}$ is actually an interesting state by itself. When $G=\ZZ_2$, after an $X$-$Z$ basis change on all edges, $\ket{\psi}$ is the so-called cluster state, a symmetry protected topological state with a 0-form and a 1-form $\ZZ_2$ symmetries \cite{Yoshida2016Higher}. Interestingly, the close relation between this cluster state and the toric code topological order has been showing up repeatedly \cite{Nat2021KWfromMeasurement,Zohar2022KWDuality,LiuJiNoninvertible}.

\section{Nonabelian Anyon Braiding}
Similar to the toric code example, intrinsic properties of density matrices restrict possible excitations above the aforementioned quantum double ground states. 
Nonetheless, we will show that \emph{nonabelian} anyon braiding can be implemented in the doubled Hilbert space. 

\begin{figure}
	\centering
	\includegraphics{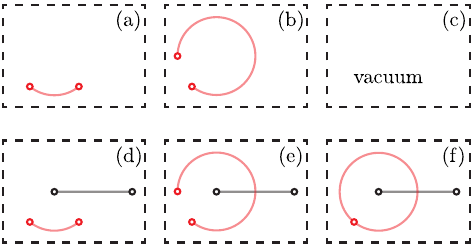}
	\caption{Schematic illustration of nonabelian braiding. Circles represent anyon excitations, and lines represent string operators used for creating anyons. The lines can be smoothly deformed without changing the underlying state as long as they do not touch other anyons. Comparing panels a-c and d-f, we see the effect of nontrivial braiding. }
	\label{fig:NonabelianSchematic}
\end{figure}

Let us begin with a nontechnical sketch of the nonabelian braiding phenomenon. In a general quantum double model, we consider a type of unitary string operators that can create anyon pairs from the vacuum, as illustrated in Fig.\,\ref{fig:NonabelianSchematic}a. If we move around one anyon from such a pair (Fig.\,\ref{fig:NonabelianSchematic}b), and finally let it approach the other anyon, then the two anyons can perfectly annihilate (Fig.\,\ref{fig:NonabelianSchematic}c), as long as the anyon trajectory does not enclose any other anyon or a hole of the space. On the other hand, if the trajectory of this anyon pair encloses another anyon, then it is possible that the anyon pair does not annihilate, but instead fuses into a new nontrivial anyon. This is illustrated in Fig.\,\ref{fig:NonabelianSchematic}d-f. 
In the following, we elaborate the full details of this phenomenon and show that it is realizable in the doubled Hilbert space. 

Before going into the doubled space, let us first try to understand an example of nonabelian braiding in the normal setting. Point-like excitations, or anyons, of a $2d$ topological order are generated by string-like operators. For our purpose here, let us consider a class of string operators $A_{\mc L}(g)$ whose action is as follows. 
\begin{align}
	&A_{\mc L}(g)\bigg|\dia{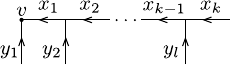}{-10}{0}\bigg\rangle\nonumber\\
	=&\bigg|\dia{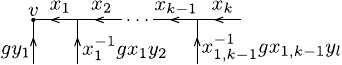}{-10}{0}\bigg\rangle. 
	\label{eq:StringOperatorDef}
\end{align}
A number of notations need explanations here: As the figure suggests, the operator $A_{\mc L}(g)$ acts within a comb-like region. $\mc L$ is an oriented string coinciding with the base of the comb (horizontal line above) and ends at the vertex $v$ (hence its orientation agrees with the horizontal arrows above). $x_{1,k-1}$ is a short-hand notation for $x_1x_2\cdots x_{k-1}$. The number of comb teeth $l$ may not equal to the string length $k$, depending on the shape of $\mc L$ and the lattice structure. We note that in the group element basis, $A_{\mc L}(g)$ is a real unitary operator. 

A remarkable property of $A_{\mc L}(g)$ is that it commutes with all $A_v$ and $B_f$ terms away from the two ends of $\mc L$. Hence, when $\mc L$ is an open string, the operator creates a pair of point-like excitations. In the special case where $\mc L$ forms a closed contractible loop, the operator $A_{\mc L}(g)$ acts trivially on all ground states of the quantum double model -- creating no excitation. Physically, this corresponds to first creating a pair of anyons and then annihilating them back to the vacuum.  

\begin{figure}
	\centering
	\includegraphics{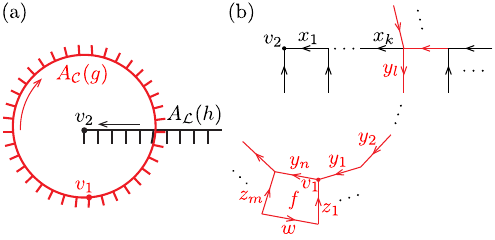}
	\caption{(a) Illustration of nonabelian braiding. The closed loop $\mc C$ is assumed to be contractible. (b) A spin configuration used for the computation of nonabelian braiding. }
	\label{fig:Braiding}
\end{figure}

To demonstrate the phenomenon of nonabelian braiding, consider two string operators $A_{\mc C}(g)$ and $A_{\mc L}(h)$ illustrated in Fig.\,\ref{fig:Braiding}a. where $\mc C$ is a contractible closed loop and $\mc L$ is open. Let $\ket{\Omega}$ be an arbitrary ground state of the quantum double model, we will show that in general $A_{\mc C}(g)A_{\mc L}(h)\ket{\Omega}\neq A_{\mc L}(h)A_{\mc C}(g)\ket{\Omega}$. Hence there is a nontrivial braiding. 
Moreover, in contrast to $A_{\mc L}(h)A_{\mc C}(g)\ket{\Omega}$ that has two anyons at the ends of $\mc L$, $A_{\mc C}(g)A_{\mc L}(h)\ket{\Omega}$ has one more anyon located at $v_1$, the base point of $\mc C$. This phenomenon is not possible for abelian braiding. 

Consider a particular spin configuration as shown in Fig.\,\ref{fig:Braiding}b, and assume the absence of flux, i.e. $B_p=1$ for all faces $p$. Let us first analyze the action of string operators on such a state: Acting $A_{\mc L}(h)$, the $y_l$ spin is mapped to $\tilde y_l:=y_lx_{1,k}^{-1}h^{-1}x_{1,k}$. Next, acting $A_{\mc C}(g)$, we have
\begin{align}
	z_1&\mapsto gz_1,\nonumber\\
	z_m&\mapsto (y_1\cdots \tilde y_l\cdots y_{n-1})^{-1}g(y_1\cdots \tilde y_l\cdots y_{n-1})z_m. 
\end{align}
Before the action of these two operators, the flux around the face $f$ vanishes: $z_1 w z_m^{-1} y_n=1$. After the action, the product of group elements around the same face becomes
\begin{align}
	&g z_1 w z_m^{-1}(y_1\cdots \tilde y_l\cdots y_{n-1})^{-1}g^{-1}(y_1\cdots \tilde y_l\cdots y_{n-1})y_n \nonumber\\
	=&g(y_1\cdots \tilde y_l\cdots y_n)^{-1}g^{-1}(y_1\cdots \tilde y_l\cdots y_n). 
\end{align}
Using $y_1y_2\cdots y_n=1$, we observe that $y_1\cdots \tilde y_l\cdots y_n$ is in the conjugacy class of $h^{-1}$, i.e. it equals to $a^{-1}h^{-1}a$ for some $a\in G$. Define $\tilde h^{-1}:=a^{-1}h^{-1}a$, then $\tilde h$ is in the conjugacy class of $h$, and the result in the previous equation simplifies to $g\tilde h g^{-1} \tilde h^{-1}$ -- the commutator between $g$ and $\tilde h$. In general, this commutator is nontrivial, leading to a violation of the $B_f$ term in the Hamiltonian. Importantly, for any other nearby face $f'\neq f$, $B_{f'}$ is preserved. This implies that the excitation at the face $f$ is a true anyon that can not be annihilated by local unitary operators, because it can be detected remotely by multiplying group elements along a large loop. The ground state $\ket{\Omega}$ is a superposition of many flux-free spin configurations, where $\tilde h$ can take all possible values among the conjugacy class of $h$. Generically, the $B_f=1$ condition will be partially or fully violated in $A_{\mc C}(g)A_{\mc L}(h)\ket{\Omega}$. As a example, consider the simplest nonabelian group $S_3$. It contains three conjugacy classes: $C_1=\{1\}$, $C_2$ consisting of three transpositions, and $C_3$ consisting of two permutations of order $3$. If we take $g\in C_2$ and $h\in C_3$, then $g$ does not commute with any possible $\tilde h$, and the resulting state $A_{\mc C}(g)A_{\mc L}(h)\ket{\Omega}$ has $B_f=0$. 

Finally, let us explain how to demonstrate nonabelian braiding in the doubled space. Recall that the string operators $A_{\mc L}(h)$ are real unitary operators, so we can directly apply them to the density matrix: 
$\rho\mapsto A_{\mc L}(h)\rho A_{\mc L}^\da(h)$, or equivalently, $\ket{\rho}\mapsto A_{\mc L}(h)\otimes A_{\mc L}(h)\ket{\rho}$. We previously mentioned that in the doubled Hilbert space, there is a picture of reducing a bilayer lattice to a single layer. Under this reduction, $A_{\mc L}(h)\otimes A_{\mc L}(h)$ is mapped to a single copy of $A_{\mc L}(h)$. We have thus understood how to apply the string operators on the doubled-space topological order. In order to observe nonabelian braiding, we need to compare the following two states (in the doubled space)
\begin{align}
	\ket{\rho_1}&=[A_{\mc C}(g)\otimes A_{\mc C}(g)][A_{\mc L}(h)\otimes A_{\mc L}(h)]\ket{\rho}, \\
	\ket{\rho_2}&=[A_{\mc L}(h)\otimes A_{\mc L}(h)][A_{\mc C}(g)\otimes A_{\mc C}(g)]\ket{\rho}, 
\end{align}
and show that $\ket{\rho_1}$ contains one more anyon excitaton. 
A possible strategy is as follows: Prepare an ancillary qubit in the $\ket{\up}$ state, and then send the full density matrix $(\ket{\up}\bra{\up})\otimes\rho_i$ ($i=1,2$) through the following quantum channel. 
\begin{align}
	\mc E_f[\sigma]&:=(Z\otimes B_f)\sigma(Z\otimes B_f)\nonumber\\
	&+[X\otimes (1-B_f)]\sigma[X\otimes (1-B_f)]. 
\end{align}
In the case of $\rho_2$, the resulting ancillary qubit will still be in the $\ket{\up}$ state. On the contrary, in the case of $\rho_1$, the ancillary qubit will have a nontrivial probability of in the $\ket{\down}$ state. 

We note that the string operator $A_{\mc L}(g)$ defined in Eq.\,\ref{eq:StringOperatorDef} can be step-by-step built up using local unitaries. To see this, define a unitary operator $E_{u_1,u_2}$ as follows.  
\begin{align}
	E_{u_1,u_2}\bigg|\dia{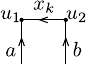}{-10}{0}\bigg\rangle=\bigg|\dia{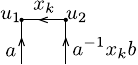}{-10}{0}\bigg\rangle. 
\end{align}
As the figure already suggests, we choose the two vertices $u_1$ and $u_2$ such that they sandwich the edge labeled by $x_k$ in Eq.\,\ref{eq:StringOperatorDef}. Then one can check that 
\begin{align}
	E_{u_1,u_2}^{-1}A_{\mc L}(g)E_{u_1,u_2}=A_{\mc L'}(g), 
	\label{eq:ElongateStringOp}
\end{align}
where $\mc L'$ is longer than $\mc L$ by one lattice unit. 

\section{Discussion}
We have shown that the preparation of many topologically ordered states are simpler in the doubled Hilbert space than the conventional Hilbert space. We have also found that nontrivial anyon braiding effects, both abelian and nonabelian, can be realized in the doubled space. What we are unable to figure out is whether such doubled-space topological orders are useful for fault-tolerant quantum computation. The intrinsic restrictions on the accessible states is a big problem and future works are needed to elucidate whether this problem sets an essential obstacle for practical applications. Nonetheless, we hope our work illustrates the potential of quantum simulation in the doubled Hilbert space and can inspire more interesting results in this direction.

\section*{Acknowledgments}
I am grateful to Yimu Bao, Yanting Cheng, Chengshu Li, and Pengfei Zhang for detailed suggestions on the writing of this manuscript. I would like to also thank Yimu Bao, Nat Tantivasadakarn, and Zijian Wang for fruitful discussions. I am supported by the Gordon and Betty Moore Foundation under Grant No.\,GBMF8690 and the National Science Foundation under Grant No.\,NSF~PHY-1748958. This work was initiated at Aspen Center for Physics, which is supported by National Science Foundation grant PHY-2210452. 
Part of this work was conducted at the Institute for Advanced Study at Tsinghua University. I would like to express my gratitude for the hospitality.



\bibliography{Bib_Refs.bib,Bib_TOQuantumSimulations}
\end{document}